\documentstyle[tighten,aps,epsfig,amssymb]{revtex} 

\begin{document} 
\title {The uphill turtle race; on short time nucleation probabilities.}
\author{Henk van Beijeren}
\address{ Institute for Theoretical Physics, University of Utrecht\\Leuvenlaan 4, 3584 CE  Utrecht, The Netherlands} 
\date{\today} 

\maketitle

\begin{abstract}
The short time behavior of nucleation probabilities is studied by representing nucleation
as a diffusion process in a potential well with escape over a barrier. If initially all growing nuclei 
start at the bottom of the well, the first nucleation time on average is larger than the inverse 
nucleation frequency. Explicit expressions are obtained for the short time probability of first nucleation.
For very short times these become independent of the shape of the potential well.
They agree well with numerical results from an exact enumeration scheme. For a large number $N$ of
growing nuclei the average first nucleation time scales as $1/\log N$ in contrast to the long-time 
nucleation frequency, which scales as $1/N$. For linear potential wells closed form expressions
are obtained for all times.
\end{abstract}


\section{Introduction}\label{sec:int}

For large systems in a metastable state the rate of nucleation of droplets of the stable phase 
is proportional to system size, in other words the average time of formation of the first 
stable nucleus in a quasi-stationary metastable system may be expected to be inversely 
proportional to its size. At first sight this may convey the impression that one might shorten 
metastable lifetimes as much as one would like, just by making systems very large, but this is 
certainly too simplistic. On a large lake at a temperature slightly below freezing, a few ice 
crystals will be formed instantly, which subsequently will continue growing. But if the 
density of these crystals is very small it will still take a very long time for the lake to 
freeze over completely. In such cases the nucleation time may be defined as the average time 
it takes an arbitrary site in the system to become included in one of the growing stable 
regions. Quantitative descriptions of this scenario have first been given by 
Kolmogorov\cite{kolm}, Johnson and Mehl\cite{johnmehl} and Avrami\cite{avrami}. The main 
resulting effect is a reduction of the effective free energy barrier for nucleation by a 
factor of $d+1$, or slightly smaller, depending on the details of the growth dynamics, in $d$ 
dimensions.

Yet, for large, but not too large systems nucleation is brought about by the first nucleation 
core reaching supercritical size, which leads to a  nucleation rate that is proportional to 
system size. The basic condition for this to occur is a time for growth of a supercritical 
nucleus throughout the system, shorter than the inverse nucleation rate. Furthermore, also in 
even larger systems the distribution of the first nucleation time may be of great interest, in 
case this event will cause an immediate dramatic change of the system. As an example of this, 
consider a condensator with plates separated by a dielectricum in a metastable phase near a 
metal-insulator transition. The formation of a supercritical nucleus of the metallic phase 
immediately will lead to discharging of the condensator. Other examples of such phenomena 
could include explosive chemical reactions starting after the nucleation event (where we may 
generalize from nucleation resulting from phase transformation to basically any process 
requiring the crossing of some free energy barrier that is large compared to $k_BT$).

Especially for these cases it is important to realize that mostly, due to the initial 
preparation of the system, the average time to the first nucleation in fact is longer than the 
inverse of the asymptotic nucleation rate. The reason is that typically metastable states are 
formed by a rapid quench from a 
stable state, in which no large clusters of the new stable phase are present. The asymptotic 
state with a constant nucleation rate, on the contrary requires a size distribution for the 
nucleating clusters
assigning non-zero probability (though small for large clusters) to clusters of any 
subcritical size.

This situation may be likened to an uphill turtle race where a large number of turtles
is released at the bottom of a wide road leading up a hill, such that each of the turtles
makes an independent random walk with a bias in the downhill direction. After a long time
the turtles that have not reached the top yet will be distributed in some characteristic way 
along the slope, with most turtles near the bottom, but also some near the top, and an arrival 
frequency $\nu$ per turtle. The average time between subsequent arrivals at the top will be 
$(n\nu)^{-1}$,
with $n$ the number of remaining turtles.
The average time of the first turtle to reach the top, however, will be much longer than
$(N\nu)^{-1}$, with $N$ the total number of turtles, due to the fact that initially all 
turtles are at the bottom. Roughly one may say the first arrival time contains a delay 
contribution, which is the time needed to set up the quasi-stationary asymptotic distribution 
starting from the initial state.

In this paper I will address the distribution of first nucleation times by approximating the 
nucleation process as a diffusion process in an abstract one-dimensional space, where the 
spatial coordinate indicates the sizes of growing nuclei. The next section gives the exact 
solution for a linear potential, corresponding to the case where the uphill road has constant 
slope. Section \ref{sec:gen} treats the case of a general monotonic potential, 
section \ref{sec:num} compares predictions to results obtained by numerically solving the 
uphill diffusion equation and the last section contains some concluding remarks.

\section{Exact solution for linear potentials}\label{sec:lin}

Mean first passage times have been studied for a long time for diffusion in a 
potential well with the possibility of escape over a potential 
barrier~\cite{nicobook}. In one dimension explicit expressions are known. 
For long times the probability of survival in the well without reaching the barrier
takes the exponential form 
\begin{equation}
S^{long}(t)=\exp[-\nu_{long}( t-t_D)],
\label{slong}
\end{equation} 
corresponding to
an escape rate
$\nu_{long}$. The delay time $t_D$ will depend on the initial distribution of the diffusor, 
but typically be much shorter than the average escape time $1/\nu_{long}$. The probability 
distribution of survival in the well for a diffusor starting at some well-defined initial 
position has been studied much less. Yet the properties of this distribution, especially for a 
starting point at the bottom of the well are of great interest in many practical situations. 
We may use again our analogy of the turtle race from the bottom to the top of the hill. If we 
have $N$ independent turtles, all starting from the same initial distribution, the probability 
distribution for the time of first passage of the top by any of them is related to the 
single-turtle survival probability as

\begin{equation} P^{arr}_N(t)=-\frac  d {d\,t}S(t)^N.  
\label{firstarr} 
\end{equation} 
If $S(t)$ were exponential for all times the distribution of first arrivals 
would be exponential likewise, with a maximum at $t=0$ and an average first 
arrival time inversely proportional to $N$. For very short times or very 
large $N$ this clearly is unrealistic; it ignores the fact that all turtles 
start at the bottom of the hill and therefore will require some minimal time 
before they can arrive at the top at all.  Obviously for short times the 
distribution of first arrivals has to be quite different from exponential.  
Van Kampen~\cite{nico} has considered the case where the turtles start 
somewhere on the middle of the slope and describes the motion by a one
dimensional diffusion equation of the form
\begin{equation} \frac {\partial \rho(x,t)}{\partial t}=\frac {\partial} 
{\partial x}\left\{ D 
\left[\frac {\partial\rho(x,t)}{\partial x}+ \frac {\partial\beta\phi(x)} 
{\partial 
x}\rho(x,t)\right]\right\}.  
\label{diffeq}
\end{equation} 
Here the diffusion constant $D$ is assumed 
constant\footnote{If $D$ depends on $x$ the equation may be transformed to a diffusion 
equation with constant $D$ by replacing $x$ by $y$ satisfying
$\frac{d\,y}{d\,x} = \left(\frac D{D(x)}\right)^{1/2}$. The potential has to be adjusted 
accordingly.}, 
$\phi(x)$
describes the external potential representing the hill, and $\beta=1/(k_B T)$,
with $T$ temperature and $k_B$ Boltzmann's constant.  Van Kampen then shows 
very elegantly that the distribution function for first arrival at $L$, 
starting from $x$, for very short times is given by 
\begin{equation}
P^{sh}_1(t)=\frac {L-x}{\sqrt{4\pi Dt^3}} \exp{-\left(\frac {(L-x)^2}{4Dt}+
\frac  {\beta  (\phi(L)-\phi(x))} 2\right)}.  
\label{pshort} 
\end{equation} 
However, his result is restricted
to really short times and it cannot be applied right away to the case where 
one starts from the origin, with a reflecting boundary imposed there.

Here I will extend his results so as to remove these limitations. The case of a strictly linear potential is
solved exactly in the present section, and in the next section short time approximations are obtained for a potential hill of general shape. 
In section \ref{sec:num} a comparison is made to numerical
solutions of the diffusion equation and it is confirmed that the approximations made in 
section \ref{sec:gen} are asymptotically correct for short enough times. 
The average time of first arrival with $N$ turtles starting from the origin is found to decrease as 
$1/\log N$ for large $N$.

To study the escape process in a potential $\phi(x)$ one may start by 
considering 
a continuous time random walk (CTRW) on a discrete set of points $1,2...L$ 
located at positions 
$x_n=n\Delta x$, with jump rates $\Gamma_{\pm}(x_n)$ for jumps to the right and 
to the left respectively, defined through
\begin{equation}
\Gamma_{\pm}(x)=\frac{\Gamma \exp[ \epsilon_{\pm}(x)]}{\exp[ \epsilon_{+}(x)]+\exp[ \epsilon_{-}(x)]},
\label{gamma}
\end{equation}
with 
\begin{equation}
\epsilon_{\pm}(x)=\frac{-\beta(\phi(x\pm \Delta x) - \phi(x))} 2.
\label{epsilon}
\end{equation} 
In a linear potential
$\epsilon$ is a constant and the forward and backward jump rates $\Gamma_+$ respectively 
$\Gamma_-$ are constants as well. As a consequence, for each CTRW 
realization
bringing a walker in $n$ steps from position $x$ to $L$ the weight equals the 
weight of the
same realization in a symmetric CTRW with the same total jump rate $\Gamma$, times
$\exp[-\beta(\phi(L)-\phi(x)]/ 2)/ \cosh^n\epsilon$.  Now 
consider the continuum 
limit where $\Delta x \to 0$
and the jump frequency $\Gamma$ is related to the diffusion coefficient through
\begin{equation}  
\Gamma (\Delta x)^2/2=D.
\label{gammavsd}
\end{equation} 
In this limit we may identify $n$ with $\Gamma t$, and the 
denominator $\cosh^n \epsilon
$ becomes $\exp\nu_{sh}t$, where I 
introduced 
\begin{equation}
\nu_{sh}\equiv\frac  1 4 (\beta\phi')^2 D. 
\label{nush}
\end{equation}
The prime denotes the derivative with respect to $x$. 
Further, the probability density of
first arrival at $L$ at time $t$ is obtained with the aid of the method of images as

\begin{eqnarray} &&P_1(x,L,t)=
\frac {L-x}{\sqrt{4\pi Dt^3}}e^{-\frac {\beta(\phi(L)-\phi(x))}2}\, 
e^{-\frac {(L-x)^2}{4Dt}}e^{-\frac  1 4 (\beta\phi')^2Dt}
\label{P1} 
\end{eqnarray}
 
This is a well-known result, see e.g.\ Ref.~\cite{grimmet}. It confirms Van 
Kampen's short time behavior, but
there is an additional damping factor which becomes important at slightly 
longer times. 

For obtaining the first arrival time distribution for escape at $x=L$ with a reflecting 
boundary at $x=0$ we can use a convolution of the return probability at the 
origin $R(t)$ with the probability density $P_1^{abs}(\Delta x,L,t)$ for first arrival
at $L$ starting from site 1, with an absorbing boundary condition at the origin. 
The latter may be calculated again by relating the CTRW in a linear potential to 
the symmetric CTRW. The method of images, now applied both at $x=0$ and 
$x=L$, yields
\begin{equation}
P_1^{abs}(\Delta x,L,t)
=4\Delta x  e^{-\frac {\beta\Delta\phi} {2}} e^{-\nu_{sh} t} \frac {\partial} 
{\partial t} 
\frac {e^{-\frac {L^2}{4Dt}}}{\sqrt{4\pi Dt}},
\label{pabs}
\end{equation}
with $\Delta\phi=\phi(L)-\phi(0)$. 
The effects of images resulting from 
repeated reflections were neglected, as these effects are exponentially small 
in the parameter $\nu_{sh}L^2/D $, which should be $\gg 1$.

By integrating this equation over time one finds that the total probability that a walk 
starting from $n=1$ will escape before returning to the origin is given by
\begin{equation}
P^{abs}=2\Delta x  e^{-\frac {\beta\Delta\phi} 2}\left(\frac{\nu_{sh}} D\right)^{1/2}  
e^{-\left(\frac{\nu_{sh}L^2} D\right)^{1/2}}.
\label{pabstot}
\end{equation}

For calculating the time dependent probability density for return to the origin 
of a walk starting at site 1
one may consider a CTRW on a semi-infinite chain in a 
linear potential, but now with transitions from site 0 to the left forbidden 
(implying that the 
total jump rate from site 0 is reduced to $\Gamma_+$). Let $X(t)$ denote the 
probability density 
for a first return at time $t$ to an initial site different from the 
origin, with the additional condition that this return is from the right. One easily 
shows~\cite{revmod} that its Laplace transform satisfies the equation
\begin{equation}
\tilde{X}(z)=\frac{\Gamma_+\Gamma_-}{(\Gamma+z)^2(1-\tilde{X}(z))}
\label{X}
\end{equation}
with the solution
\begin{equation}
\tilde{X}(z)=\frac{z+\Gamma-\sqrt{z^2+ 2\Gamma 
z+\Gamma^2\epsilon^2}}{2(z+\Gamma)},
\label{solX}
\end{equation}
where  Eq.~\ref{gamma} was used.  The Laplace transform for the distribution of 
return times at the origin
now is obtained by summing  a geometric series over $n$ returns as
\begin{eqnarray}
\tilde{R}(z)&=&\left[1-\frac 1 {(z+\Gamma_+)}(z+\Gamma)\tilde{X}(z)\right]^{-1} 
\nonumber\\
          &=&\frac{z+\Gamma_+}{\frac z 2 +\Gamma_+ -\Gamma_ - +\frac 1 2 
\sqrt{z^2+2\Gamma 
z+\epsilon^2\Gamma^2}}.
\label{R}
\end{eqnarray}
Next, note that, for fixed $D$, $\Gamma$ scales as $1/\epsilon^2$ as $\epsilon \to 0$. 
Therefore, dividing both the 
numerator and the denominator of the expression above by $\Gamma$, keeping the 
dominant terms in $\epsilon$ and using Eqs.\ (\ref{gamma}), (\ref{epsilon}) and (\ref{nush}),
we end up with
\begin{eqnarray}
\tilde{R}^{lin}(z)&=& \frac{1}{\epsilon(\sqrt{\frac {z} {\nu_{0}}+1}-1)},
\label{R2}
\end{eqnarray}
where the superscript denotes the solution 
for a linear potential. In the present case $\nu_0$ is identical to $\nu_{sh}$. 
An inverse 
Laplace transform yields
\begin{equation}
R^{lin}(t)=\frac{\beta \phi'\Gamma \Delta x} 4 
\left(1+\frac{e^{-\nu_{0}t}}{\sqrt{\pi\nu_{0}t}} + 
erf\sqrt{\nu_{0} t}\right),
\label{rlin}
\end{equation}
with $erf(x)=\frac 2 {\sqrt{\pi}} \int^x_0
d\,y \exp(-y^2)$.  Now the first arrival distribution may 
be obtained from
\begin{equation}
P^{arr}_1(t)=\int_0^t d\,\tau R(\tau)P_1^{abs}(\Delta x,L,t-\tau)
\label{arr}
\end{equation}
For the linear potential all integrations may be done in closed form with the 
result
\begin{eqnarray}
P^{arr}_1(t)=&&e^{-\beta \Delta\phi}{\Huge\{}\frac{ e^{-(\frac L 
{\sqrt{4Dt}}-\sqrt{\nu_0 t})^2}}{\sqrt{\pi D 
t}}\left(\beta\phi'D+\frac L t\right) \label{arr2} \\
&&+2\nu_0\left[1-erf(\frac L {\sqrt{4Dt}}-\sqrt{\nu_0 t}). 
\right] {\Huge\}} \nonumber
\end{eqnarray}
For very short times this agrees with Van Kampen's expression, Eq.~(\ref{pshort}).

\section{General potentials}\label{sec:gen}
For potential hills of general shape Eq.\ (\ref{arr}) of course remains valid, but
we do not have explicit solutions any more. To assess the short time behavior of the arrival 
time distribution  we need short time approximations for the the return time distribution
$R(t)$ and the first arrival distribution with absorbing boundary conditions $P_1^{abs}$.

Let us restrict ourselves to
cases in which the hill is high, i.e.\ $\beta \Delta \phi \gg 1$, the bottom of 
the hill is at the origin 
and the top
at $x=L$, there are no intermediate maxima and minima at almost the same height 
as the top or bottom,
and the shape of the hill near bottom and top is smooth over length scales on 
which the potential variations 
are of order $k_B T$. Under these conditions there are three
well-separated time scales. A short timescale is given by $t_{sh}= {L_0^2} /D$, with $L_0$ a 
characteristic distance 
from the origin where the potential has increased by an amount of order $k_B T$. On this time scale
an initial distribution localized near 
the origin approaches an 
equilibrium-like distribution over a potential range of a few $k_BT$ around the origin. This is the 
range within 
which the major part of all turtles will be found at any time. An intermediate time
$t_{med}$, is set by the average time a turtle needs to get from bottom to top, in case it 
does not return to the bottom. This is the time scale required to establish the full 
metastable distribution. The longest time scale is 
$t_{esc}=\nu_{esc}^{-1}$, the average escape time or arrival time. For the 
linear potential
one may choose 
\begin{eqnarray}
t_{sh}^{lin}&=& (4\nu_{sh})^{-1}=1/((\beta\phi')^2 D)=L^2/((\beta\Delta 
\phi)^2D), \nonumber\\
t_{med}^{lin}&=&L^2/(\beta\Delta \phi D), \nonumber\\
t_{esc}^{lin}&=&t_{sh}^{lin}\exp(-\beta \Delta\phi).
\label{tshlin}
\end{eqnarray}
Under the given assumptions 
indeed all three scales are well separated, though the separation between 
$t_{sh}^{lin}$ and $t_{med}^{lin}$
is much smaller than that between  $t_{med}^{lin}$ and $t_{esc}^{lin}$.

Now one may formulate short time approximations for more general hill shapes.
Assign to each random walk realization the Hamiltonian
\begin{equation}
H(\{x_i\})= \sum_i\left\{\frac{\phi(x_{i+1})-\phi(x_i)} 2+\frac 1 \beta 
\log\left(\frac{\exp[ \epsilon_{+}(x_i)]+\exp[ \epsilon_{-}(x_i)]} 2\right)
\label{ham}\right\},
\end{equation}
with $i$ running over all steps of the walk and $x_i$ the position before the 
$i+1^{th}$ step. The probability for moving from $x_0$ to $x_t$ in a time t under 
specific boundary conditions $BC$ then may be obtained as
\begin{equation}
P(x_0,x_t|BC)=\langle exp^{-\beta H} \rangle_{x_0,x_t,BC}\,P_0(x_0,x_t|BC),
\label{trans}
\end{equation}
with $P_0(x_0,x_t|BC)$ the corresponding probability for the unbiased random walk.
The average $\langle\rangle$ runs over all unbiased continuous 
time random walks,
properly weighted, starting at $x_0$, ending at $x_t$ and satisfying the 
required boundary conditions. For short times this average may be replaced by 
the Rosenstock approximation~\cite{rosenstock}
\begin{equation}
P(x_0,x_t|BC)= e^{-\beta\langle H\rangle_{x_0,x_t,BC}} P_0(x_0,x_t|BC).
\label{rosen}
\end{equation}
In the continuum limit this reduces to
\begin{equation}
P(x_0,x_t|BC)=\exp-\frac{\beta(\phi(x_t)-\phi(x_0))+\left\langle(\beta\phi')^2 
/2+\beta \phi''\right\rangle D t} 2 \,P_0(x_0,x_t|BC),
\label{roscont}
\end{equation}
where the subscripts on the random walk average were omitted. 
Eq.~(\ref{roscont}) is especially useful as an approximation
for $P_1^{abs}(\Delta x,L,t)$. For short times unbiased walks from $\Delta x$ to 
$L$ with absorbing boundary conditions at the origin and at $x=L$ on average spend 
equal time in equal intervals, except for very small neighborhoods of the end 
points, where the average time spent is smaller due to the absorbing boundaries. 
Therefore the average $\langle\frac{(\beta\phi')^2} 4+\frac{\beta \phi''} 2  \rangle$ may be replaced by 
a spatial average over the interval $(0,L)$ and one obtains the approximation
\begin{equation}
P_{1sh}^{abs}(\Delta x,L,t)
=4\Delta x  e^{-\frac {\beta\Delta\phi} {2}} e^{-\nu_{abs}t} \frac {\partial} 
{\partial t} 
\frac {e^{-\frac {L^2}{4Dt}}}{\sqrt{4\pi Dt}},
\label{p1sh}
\end{equation}
with $\nu_{abs}$ the spatial average of $(\frac{(\beta\phi')^2} 4 +\frac{\beta \phi''} 2 )D$.

For the return probability to the origin the approximation (\ref{roscont}) in 
principle could be used as well, but in this case it gives rise to somewhat 
cumbersome integrals involving error functions. And in fact we don't really need this: for short times in typical cases the diffusion effectively takes place
near the origin in either a linear or a quadratic potential, so one may approximate the return probability by the explicit expressions 
for these potentials. In the case of a linear potential this becomes Eq.\ (\ref{rlin}), with 
\begin{equation}
\nu_0=\frac{(\beta \phi'(0))^2}{4},
\label{nu0}
\end{equation}
and $\phi'$ replaced by $\phi'(0)$ likewise.
For potentials that are quadratic near the origin the return probability may be obtained from the Green function $G(x,x_0,t)$ for diffusion in a quadratic well \cite{nicogreen} as
\begin{eqnarray}
R^{qu}(t)&=& \Gamma  \Delta x G( 0,0,t) \nonumber\\
&=&\frac{\Gamma\Delta x\kappa(0)}{2\sqrt{(1-\exp[{-2\beta \phi''(0) Dt}])}},
\label{RQ}
\end{eqnarray}
with $\kappa(x)$ defined as~\footnote{This may be generalized to the case of a potential of form 
$\phi(x)=\phi(0)+cx^{\alpha}$, 
respectively $\phi(x)=\phi(L)-c(L-x)^{\alpha}$. In this case one obtains 
$\kappa=(\beta c)^{-1/\alpha}\Gamma(\frac {\alpha+1} \alpha)$.} $\kappa(x)=\sqrt{(2\beta|\phi''(x)|)/\pi}$.
In either case the short time behavior of the arrival time distribution is obtained according to Eq.\ (\ref{arr}) as the convolution of the return probability with the arrival probability with absorbing boundary conditions.
For times $\ll t_{sh}$ the exponential damping factors $\exp -\nu_0 t$ and $\exp -\nu_{abs}t$ may be ignored and one finds that the arrival probability
asymptotically behaves as
\begin{equation}
P^{arr}_{sh}(t)= \frac{L e^{-\frac{\beta\Delta\phi} 2} e^{-\frac {L^2}
{4Dt}}}{\sqrt{\pi D t^3}},
\label{parrsh}
\end{equation}
irrespective of the shape of the potential.

To find the average first arrival time for very large numbers of turtles, 
notice that for times
$\ll t_0$ the survival probability for a single turtle may be obtained from 
Eqs.\ (\ref{arr}), (\ref{p1sh}) and (\ref{parrsh}) as
\begin{equation}
S(t)=1-\int^t_0d\,\tau P^{arr}_{sh}(t)\approx 1-4\sqrt{\frac{Dt}{\pi L^2}}e^{-\frac{\beta\Delta\phi} 
2}e^{-\frac{L^2}{4Dt}}.
\label{SN}
\end{equation}
The mean first arrival time for the case of $N$ 
turtles may be found by setting the term subtracted from unity equal to $1/N$. This leads to
\begin{equation}
\bar{t}_{esc}(N)=\frac {L^2} {4D\left(\log N -\frac{\beta\Delta\phi} 2 - \log(\frac L 
4 \sqrt{\frac{\pi}{D\bar{t}_{esc}(N)}})\right)}.
\label{tescN}
\end{equation}
From this one rapidly sees that for $N  \gg  \exp(\beta\Delta\phi /2) $ the first arrival time 
approaches zero as $1/\log N$. This is much slower indeed than the 
$1/N$ behavior one would find for smaller values of $N$, such that typically a 
quasi-stationary distribution over the full slope is reached well before the first turtle escapes.

Eqs.\ (\ref{firstarr}), (\ref{parrsh}) and (\ref{SN}) may also be used to consider fluctuations in ${t}_{esc}(N)$.
One readily finds that
\begin{equation}
\left(\frac{{t}_{esc}(N)-\bar{t}_{esc}(N)}{\bar{t}_{esc}(N)}\right)^2\sim \frac 1{\log^2 N}.
\end{equation}
So the distribution of ${t}_{esc}(N)$ becomes sharp for very large $N$.
This is in marked contrast to the case of a Poisson distribution, where relative fluctuations
are independent of $N$.

\section{Numerical results}\label{sec:num}

Numerical results were obtained by solving discrete time random walks on a lattice of $L$ sites, in a number of different potentials. At each time step a fixed fraction $\Gamma$ of the walkers are moved to
their neighboring sites, with jump probabilities satisfying Eq.\ (\ref{gamma}).
Walkers reaching the top of the hill are taken out of the system. This is done most efficiently in an exact enumeration scheme, where 
one starts from an initial density distribution, typically concentrated at the origin, and evolves this distribution in discrete time 
in accordance with the jump probabilities. In this way it is possible to capture also the very small arrival probabilities at short 
times. 

\begin{figure}[h]
\center{\epsfig{file=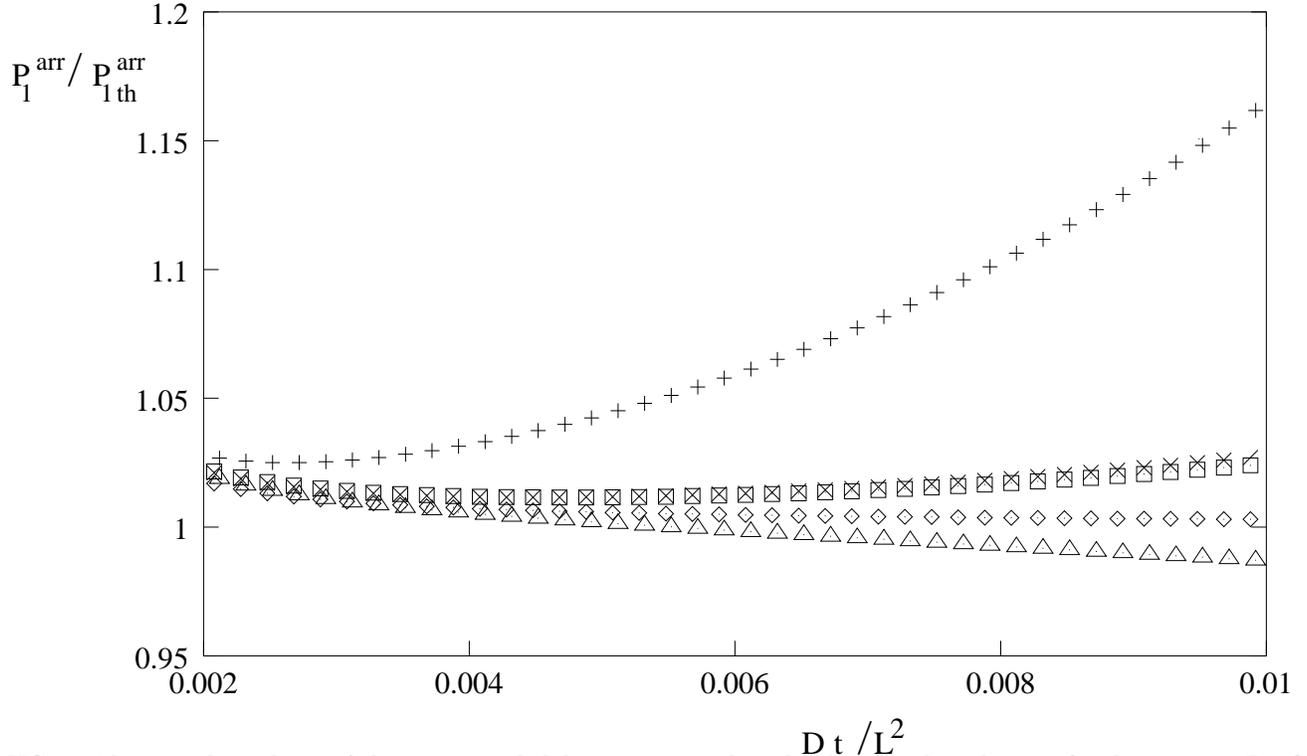,height=10cm}}
\caption{The time dependence of the escape probability is compared to the theoretical predictions for short times. For five different 
shapes of the potential hill the ratio of the numerical solution of the diffusion equation to the theoretical
expression is plotted as function of $Dt/L^2$. The exact enumeration results for the linear potential 
(diamonds) are compared to the 
exact solution, Eq.\ (\ref{arr2}) and those for the quadratic (crosses), the inverse qudratic (squares), the cubic (pluses) and the 
cosine 
(triangles) potential
to the numerical solution of Eqs.\ (\ref{arr}) together with (\ref{p1sh}) and (\ref{rlin}) or (\ref{RQ}) }
\label{fig}
\end{figure}

In all the calculations reported here $L=10,000$, $\Gamma=0.04$ and the potential difference between bottom and top of the hill is $\Delta\phi=20k_BT$. The potentials considered were of the forms
\[
\begin {array}{ll}
\phi(x)=\Delta\phi\, \frac x L&{linear,}\\
\phi(x)=\Delta\phi\, (\frac x L)^2 &{quadratic,}\\
\phi(x)=\Delta\phi\, \left(\frac{2 x }L-\left({\frac x L}\right)^2\right)&{inverse\  quadratic,}\\
\phi(x)=\Delta\phi\, \left(4\,(\frac x L -\frac 1 2 )^3+ \frac 1 2 \right)&{cubic,}\\
\phi(x)=\frac 1 2 \Delta\phi\, (1-\cos \frac{\pi x} L) &{cosine.} 
\end{array}
\]
Figure \ref{fig} shows a comparison of the first arrival probabilities resulting from the exact enumeration scheme to the predictions 
of Eq.\ 
(\ref{arr}), combined with (\ref{p1sh}) and (\ref{rlin}) or (\ref{RQ}). The linear potential, for which we have the exact result (\ref{arr2}), provides a check on the accuracy of the discretized dynamics as an approximation for the diffusion equation\footnote{Here one should keep in mind that in actual applications often the diffusion equation 
is obtained as a continuum approximation for dynamics that are in reality discrete in the spatial coordinates.}. Figure \ref{fig}
shows for the linear potential the ratio of the exact enumeration results to those of Eq.\ (\ref{arr2}),
as function of the dimensionless time $\tau \equiv \Gamma n/2L^2$, with $n$ the discrete time in the enumeration scheme. On this scale the relaxation time $\nu_0^{-1}$ corresponds to $\tau=0.01$. One sees that the discretization effects remain limited to less than 2.5\%, on the shortest time scales yielding
an arrival probability different from zero within the computer accuracy. They decay to less than 1\% for larger $\tau$. Instead of Eq.\ (\ref{arr2}) one may
also use (\ref{arr}), combined with (\ref{p1sh}) and (\ref{rlin}). This everywhere yields slightly larger values, but the difference never exceeds
0.4\%.
For all the other potentials the results for times up to $t_{br}$, with $t_{br}=$ min$(\nu_0^{-1},\nu_{abs}^{-1}),$ remain within a deviation of 
3\% of the analytic approximation. The largest deviations occur for the cubic potential, which indeed of all potentials considered has the shortest $t_{br}$ (namely $t_{br}=0.0011\cdots$). Part of the deviations
always are due to discretization errors, as seen already for the linear potential. But given that the numbers divided upon each other to obtain these curves, easily vary over more than a hundred orders of magnitude within the time range considered, errors of a few percent may be considered quite a good result.

\section{Discussion\label{sec:disc}}
In this paper I obtained the short time behavior of the first arrival probability at the top of a 
potential hill for a diffusion process or random walk starting at the bottom. It is strongly suppressed during an initial 
time interval on the order of the diffusion time from bottom to top in the absence 
of a potential. In the continuum diffusion description it never becomes strictly zero for positive times, but it 
approaches zero 
extremely rapidly as time goes to zero.  The arrival probability for very short times becomes fully independent of the shape of the 
potential, but the time range over which this holds becomes shorter as the potential gets higher and steeper near the origin.

For slightly longer times the inclusion of exponential damping factors in the expressions for the first arrival 
probability
does become important. E.g.\ ignoring the second derivative term in the expression for
$\nu_{abs}$ below Eq.\ (ref{p1sh}) leads to a deviation of roughly 20\% at $\tau=0.01$ for the quadratic and the inverse 
quadratic potential, whereas the deviations with the full expression are only about 1\%.

It is obvious that the delay time $t_D$ introduced in Eq.\ (\ref{slong}) has to be of the order $t_{med}$, so the factor 
$\exp (\nu_{long}t_D)$ is very close to unity. Explicit expressions for $t_D$ may be obtained from the projection of the 
initial 
distribution on the most slowly decaying eigenfunction of the diffusion equation (\ref{diffeq}) with escape at $x=L$. Most notable is 
that $t_D$ in essence is independent of the precise form of the initial distribution, as long as this remains localized within a region 
of width $L_0$ around the origin, where the value of the potential remains less than a few $k_BT$ above that in 
the origin.

The average first arrival time for a very large number of independent random walkers, all starting at or near the bottom, does not 
scale as 
the inverse of the number of walkers $N$, as one might expect on the basis of Poisson statistics, but rather as $1/\log\ N$. This may have important consequences in large metastable systems, in which the first nucleation of a stable droplet has an immediate large effect on the whole system.

It is an interesting question how accurately such systems may be described by
a simple model of noninteracting random walks in one dimension. A priori it is not clear that the nucleating droplets are characterized 
sufficiently by a single parameter
giving  their size, ignoring all details about their shapes. Interactions between droplets may play a role, especially when their 
density becomes larger. And, especially in the presence of conservation laws there may be memory effects that make a simple random 
walk picture inadequate. Presently these questions are under investigation both numerically and analytically and we expect to report on
them soon~\cite{tbp}.

\acknowledgements
With great pleasure I dedicate this paper to Michael Fisher, who has been an enlightening guide showing us new ways to go for so  many decades now. I wish him an equally good continuation of his activities after his seventieth birthday.

I thank Gerard Barkema and Reinier Bikker for very helpful discussions and assistance with computer manipulations.

Support by the Statistical physics program of FOM is gratefully acknowledged.

\end{document}